\documentclass[
]{aa}  

\usepackage{supertabular}
\usepackage{graphicx,natbib}
\usepackage[varg]{txfonts}
\usepackage{dcolumn}
\usepackage{lscape}
\usepackage{amsmath}
\usepackage{booktabs}

\usepackage{color}
\definecolor{dblue}{rgb}{0.0, 0.0, 0.65}
\usepackage[
    pdfauthor={Steve Ertel}, 
    pdfsubject={The multi-component debris disk around HIP17439 as revealed by Herschel/DUNES},
    pdfmenubar=true,
    colorlinks=true,
    linkcolor=dblue,
    urlcolor=dblue,
    citecolor=dblue
]{hyperref}

\newcommand{\um}{\,\mu\textrm{m}}

\newcommand{\Spitzerr}{\textit{Spitzer}\ }                 
\newcommand{\Spitzer}{\textit{Spitzer}}                    
\newcommand{\Herschell}{\textit{Herschel}\ }
\newcommand{\Herschel}{\textit{Herschel}}

\newcommand{\AU}{\,\textrm{au}}
\newcommand{\Gyr}{\,\textrm{Gyr}}

\begin{document}
   \title{Potential multi-component structure of the debris disk around HIP\,17439 revealed by Herschel/DUNES\thanks{\Herschell is an ESA space observatory with science instruments provided by European-led Principal Investigator consortia and with important participation from NASA.}}


   \author{S. Ertel\inst{1}\thanks{The author recently changed his affiliation. New affiliation is: \mbox{European Southern Observatory}, Alonso de Cordova 3107, Vitacura, Casilla 19001, Santiago, Chile}
        \and
          J.~P. Marshall\inst{2}
        \and
          J.-C. Augereau\inst{1}
        \and
          A.~V. Krivov\inst{3}
        \and
          T. L\"ohne\inst{3}
        \and
          C. Eiroa\inst{2}
        \and
          A. Mora\inst{4}
        \and
          C. del Burgo\inst{5}
        \and
          B. Montesinos\inst{6}
        \and
          G. Bryden\inst{7}
        \and
          W. Danchi\inst{8}
        \and
          F. Kirchschlager\inst{9}
        \and  
          R. Liseau\inst{10}
        \and
          J. Maldonado\inst{2}
        \and
          G.~L. Pilbratt\inst{11}
        \and
          Ch. Sch\"uppler\inst{3}
        \and
          Ph. Th\'ebault\inst{12}
        \and
          G.~J. White\inst{13,14}
        \and
          S. Wolf\inst{9}
   }
   \institute{UJF-Grenoble 1 / CNRS-INSU, Institut de Plan\'etologie et d'Astrophysique de Grenoble (IPAG) UMR 5274, Grenoble, F-38041, France \\
          \email{steve.ertel@obs.ujf-grenoble.fr} 
        \and
          Dpt. de F\'\i sica Te\'orica, Facultad de Ciencias, Universidad Aut\'onoma de Madrid, Cantoblanco, 28049 Madrid, Spain
        \and
          Astrophysikalisches Institut und Universit{\"a}tssternwarte, Friedrich-Schiller-Universit{\"a}t, Schillerg{\"a}{\ss}chen 2-3, 07745 Jena, Germany
        \and
          ESA-ESAC Gaia SOC. P.O. Box 78, 28691 Villanueva de la Ca{\~n}ada, Madrid, Spain
        \and
          Instituto Nacional de Astrof\'\i sica, \'Optica y Electr\'onica, Luis Enrique Erro 1, Sta. Ma. Tonantzintla, Puebla, Mexico
        \and
          European Space Observatory, Alonso de Cordova 3107, Vitacura Casilla 19001, Santiago 19, Chile
        \and
          Jet Propulsion Laboratory, California Institute of Technology, Pasadena, CA 91109, USA
Spain.
        \and
          NASA Goddard Space Flight Center, Exoplanets and Stellar Astrophysics, Code\,667, Greenbelt, MD\,20771, USA
        \and
          Institut f\"ur Theoretische Physik und Astrophysik, Christian-Albrechts-Universit\"at zu Kiel, Leibnizstra{\ss}e 15, 24098 Kiel, Germany
        \and
          Onsala Space Observatory, Chalmers University of Technology, 439 92 Onsala, Sweden
        \and
          ESA Astrophysics \& Fundamental Physics Missions Division, ESTEC/SRE-SA,  Keplerlaan 1, 2201 AZ Noordwijk, The Netherlands
        \and
          Observatoire de Paris, Section de Meudon 5, place Jules Janssen, 92195 MEUDON Cedex, Laboratoire d'\'etudes spatiales et d'instrumentation en astrophysique, France
        \and
          Department of Physics and Astrophysics, Open University, Walton Hall, Milton Keynes MK7 6AA, UK
        \and
          Rutherford Appleton Laboratory, Chilton OX11 0QX, UK
   }

   \date{}

 
  \abstract
   {The dust observed in debris disks is produced through collisions of larger bodies left over from the planet/planetesimal formation process. Spatially resolving these disks permits to constrain their architecture and thus that of the underlying planetary/planetesimal system.}
   {Our \Herschell Open Time Key Programme DUNES aims at detecting and characterizing debris disks around nearby, sun-like stars. In addition to the statistical analysis of the data, the detailed study of single objects through spatially resolving the disk and detailed modeling of the data is a main goal of the project.}
   {We obtained the first observations spatially resolving the debris disk around the sun-like star HIP\,17439 (HD\,23484) using the instruments PACS and SPIRE on board the \Herschell Space Observatory. Simultaneous multi-wavelength modeling of these data together with ancillary data from the literature is presented.}
   {A standard single component disk model fails to reproduce the major axis radial profiles at $70\um$, $100\um$, and $160\um$ simultaneously. Moreover, the best-fit parameters derived from such a model suggest a very broad disk extending from few au up to few hundreds of au from the star with a nearly constant surface density which seems physically unlikely. However, the constraints from both the data and our limited theoretical investigation are not strong enough to completely rule out this model. An alternative, more plausible, and better fitting model of the system consists of two rings of dust at approx.\ 30\,au and 90\,au, respectively, while the constraints on the parameters of this model are weak due to its complexity and intrinsic degeneracies.}
   {The disk is probably composed of at least two components with different spatial locations (but not necessarily detached), while a single, broad disk is possible, but less likely. The two spatially well-separated rings of dust in our best-fit model suggest the presence of at least one high mass planet or several low-mass planets clearing the region between the two rings from planetesimals and dust.}


   \maketitle
\section{Introduction}
\label{sect_intro}
Circumstellar debris disks around main-sequence stars are most usually observed as excesses above the stellar photospheric emission at far-infrared wavelengths due to dust emission. The dust in these disks is thought to be produced by collisions of planetesimals left over from the planet formation process in the system (see \citealt{kri10} for a recent review). In addition to the stellar gravity and radiation, the presence of planets in the system may significantly sculpt the distribution of the dust and planetesimals through their gravitational interaction \citep{dom03,ken04,wya08,ert12b}. This can result in different features of the disk such as clumpy structures and multiple rings depending on the configuration and history (e.g., migration) of the underlying planetary system \citep{wya08,ert12b}. Indeed, candidates for such complex architectures have been observed in several systems (e.g., $\epsilon$\,Eridani: \citealt{bac09}; Fomalhaut: \citealt{ack12}).

HIP\,17439 (HD\,23484) is a sun-like (sp.\ type K2\,V, \citealt{tor06,gra06}), nearby ($d = 16.0\,\textrm{pc}$, \citealt{vanle07}) main-sequence star with an estimated age of 760\,Myr \citep{mal10}. The first conclusive detection of far-infrared excess was reported by \citet{koe10} based on \Spitzer/MIPS observations. Earlier detections of the excess were ascribed to contamination by galactic cirrus \citep{moo06,sil00}. In this paper, we present far-infrared \Herschell \citep{pil10} PACS \citep{pog10} and SPIRE \citep{gri10, swi10} images of the debris disk obtained in the context of our \Herschell Open Time Key Programme DUNES \citep[DUst around NEarby Stars,][]{eir10, eir13}. These data are complemented by a re-reduction of the available \Spitzerr data (MIPS and IRS) as well as by other data from the literature. We present simultaneous multi-wavelength modeling of the debris disk considering all available, relevant data in order to constrain the architecture of the dust and planetesimal disk beyond the possibilities of single data sets. This way, we put constraints on the underlying planetary/planetesimal system potentially responsible for the complex disk structure.

We present the available data in Sect.~\ref{sect_data}, including predictions of the stellar photospheric flux and a re-reduction of the \Spitzerr data being the most relevant, complementary data to the new \Herschell observations. In this section, we also describe the new \Herschell data summarizing the observations, the data reduction, results, and present a basic analysis of the \Herschell images (image deconvolution and radial profile extraction). In Sect.~\ref{sect_mod}, analytical, simultaneous model fitting to all available data is presented including a detailed interpretation of the results. A summary and conclusions are given in Sect.~\ref{sect_conc}.

\section{Observations and data analysis}
\label{sect_data}

In this section, we present the observational data of the HIP\,17439 system. This includes the characterization of the host star and the fitting of a stellar photosphere model, a re-reduction of the available Spitzer data, as well as a detailed description of our new \Herschell observations, and the subsequent data reduction and analysis. This work is done analogous to the whole survey sample. For more details on the general observing strategy and data reduction of the DUNES survey see \citet{eir13}.

\subsection{Stellar parameters and photosphere estimation}

Table~\ref{star_params} provides fundamental stellar parameters of HIP\,17439. The bolometric luminosity has been estimated from the absolute $V$ magnitude and the bolometric correction from \cite{flo96}. Effective temperature, surface gravity and metallicity are mean values of spectroscopic estimates by \cite{val05} and \cite{san04}. The projected rotational velocity is taken from \cite{tor06}. HIP\,17439 is an active star with a Ca~{\sc ii} $\log R'_{\rm HK}$ index of -4.534 \citep{gra06}. The observed ROSAT X-Ray luminosity $L_X/L_{bol}$, is also given. The age of the star estimated from the  $\log R'_{\rm HK}$ index \citep{mam08} is $0.8\Gyr$ while from the X-Ray luminosity following \cite{gar10} it is $0.9\Gyr$. A recent estimate using evolutionary tracks gives $3.7\Gyr$ \citep{fer11}, although we note that isochrones are highly degenerate for stars located right on the main sequence and are very sensitive to metallicity \citep{hol09}. Thus, we favor the consistent values estimated from the X-ray luminosity and activity index. The mass of the star using the gravity and stellar radius is 0.63\,$M_\odot$, while \cite{fer11} obtain 0.83\,$M_\odot$. 

The stellar photosphere contribution to the total flux was calculated from a synthetic stellar atmosphere model interpolated from the PHOENIX/GAIA grid \citep{bro05}. After a comparison of possible normalization methods (e.g. to all near-infrared data, WISE-only or the \Spitzer/IRS spectrum) the stellar models were scaled to the combined near-infrared and WISE band 3 fluxes following \citet{ber04}, as being the optimum solution. WISE band 2 was omitted due to saturation and WISE band 4 is in a region of the SED already exhibiting excess emission in the \Spitzer/IRS spectrum.

\begin{table}
\caption{Physical properties of HIP\,17439, cf. discussion in Section 2.1.}
\label{star_params}
\begin{tabular*}{\linewidth}{l@{\extracolsep{\fill}}l}
\toprule
Parameter   &  Value  \\
\midrule
Distance                                      & $16.0 \pm 0.1\,\textrm{pc}$ \\
Spectral type, luminosity class               & K0\,V $\dots$ K2\,V \\
$V$, $B-V$                                    & 6.98, 0.88~mag\\
Abs. magnitude $M_V$, Bol. Correction         & 5.97, -0.24 \\
Bolometric luminosity, $L_{\star}$            & 0.402~L$_\odot$ \\ 
Effective temperature                         & 5166~K \\
Surface gravity, $\log g$                     & 4.44  \\
Radius, $R_{\star}$                           & $0.80 R_\odot$ \\
Metallicity, $[\rm{Fe/H}]$                    & 0.05 \\
Rotational velocity, $\upsilon \sin i$        & 2.9~kms$^{-1}$ \\
Activity, $\log R'_{\rm{HK}}$                 & -4.534 \\
X-Ray luminosity, $\log~L_{\rm{X}}/L_{\star}$ & -4.9 \\
Mass, $M_{\star}$                             & $0.63-0.83 M_\odot$ \\
Age, $\log R'_{\rm{HK}}$                      & $0.76\Gyr$\\
Age, $\log~L_{\rm{X}}/L_{\star}$              & $0.93\Gyr$\\
Age, isochrones                               & $3.7\Gyr$\\
\bottomrule
\end{tabular*}
\end{table}

\subsection{\Spitzerr observations and photometry}

\begin{figure*}
\centering
\includegraphics[angle=0,width=1\linewidth]{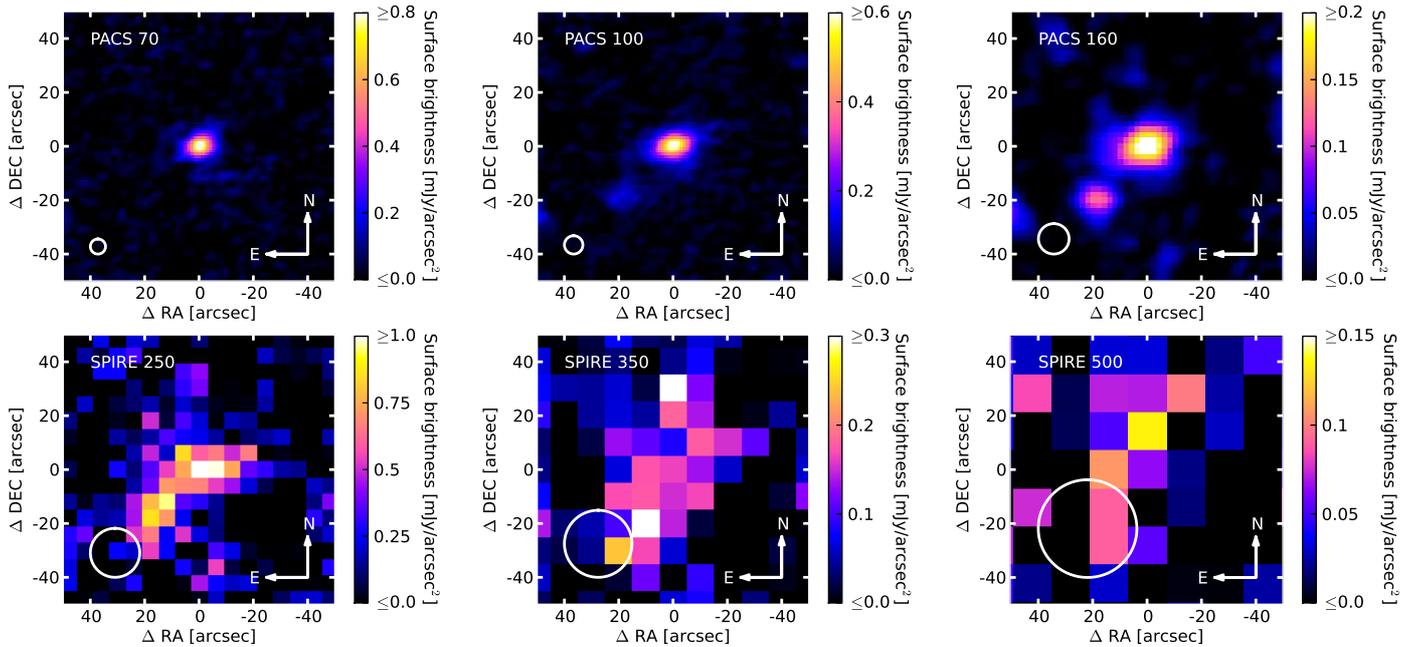}
\caption{\Herschel/PACS and \Herschel/SPIRE observations of HIP\,17439. The band is noted in the upper left corner of each panel. The white circle in the lower left of each panel illustrates the FWHM of the theoretical, circular PSF.}
\label{fig_herschelobs}
\end{figure*}

Our data analysis for the \Spitzer/MIPS images (program ID: 30490, PI: D.~W.~Koerner, \citealt{koe10}) is similar to that described in \citet{bry09}. At $24\um$, images are created from the raw data using software developed by the MIPS instrument team \citep{gor05}, with image flats chosen as a function of scan mirror position to correct for dust spots and with individual frames normalized to remove large scale gradients \citep{eng07}. At $70\um$, images are also processed with the MIPS instrument team pipeline, including added corrections for time-dependent transients \citep{gor07}. Aperture photometry at $24\um$ and $70\um$ is performed as in \citet{bei05a} with aperture radii of $15\farcs3$ and $14\farcs8$, background annuli of $30\farcs6-43\farcs4$ and $39\farcs4-78\farcs8$, and aperture corrections of 1.15 and 1.79 at $24\um$ and $70\um$, respectively. The $24\um$ centroid positions, which are consistent with the telescope pointing accuracy of $<$~$1\arcsec$ \citep{wer04}, are used as the target coordinates for both wavelengths. HIP\,17439 is observed at $24\um$ with high S/N; uncertainty at that wavelength is dominated by systematics at the level of $\sim$2\% for overall calibration and $<$~$1\%$ for repeatability \citep{eng07}. The $70\um$ uncertainties are calculated from direct measurement of the sky background variation in each field, using the same apertures and corrections as for the photometry. A calibration uncertainty of 5\% and a repeatability uncertainty of 4.5\% \citep{gor07} are also included. The different contributions to the uncertainties are added quadratically. Color corrections of 1.12 and 0.893 at $24\um$ and $70\um$, respectively, have been applied.

The \Spitzer/IRS \citep{hou04} spectrum of HIP\,17439 (program ID: 50150, PI: G.~Rieke) spans the two long wavelength, low resolution modules covering $14\um-40\um$. The spectrum presented here was taken from the CASSIS\footnote{The Cornell Atlas of Spitzer/IRS Sources (CASSIS) is a product of the Infrared Science Center at Cornell University, supported by NASA and JPL.} archive \citep{lebout11}. Data are taken with the star at two locations along the slit to permit background subtraction. The final spectrum is the average of the two slit positions, whilst the uncertainty is estimated from the difference between the value at each wavelength. The slit width ($\sim$$3\farcs6$) is greater than the \Spitzerr pointing uncertainty of $<$~$1\arcsec$, so flux loss outside the slit is minimal and no scaling of the individual modules is required \citep{wer04}. The CASSIS spectrum lies systematically above the model photosphere and we have therefore scaled the IRS spectrum by a factor of 0.93 to fit the predicted photosphere model at $14\um$. After scaling, fluxes in bands centered at $17\um$ and $32\um$ were calculated to look for evidence of warm excess and the rising excess from cold dust grains. The fluxes were measured in each band by summing the error-weighted fluxes of all points on the spectrum lying between $15\um-19\um$ and $30\um-34\um$ following \cite{che06,che07}.

\subsection{\Herschell observations and photometry}

HIP\,17439 was observed with both PACS and SPIRE instruments. The resulting images at $70\um$ to $500\um$ are shown in Fig.~\ref{fig_herschelobs} along with a circle illustrating the angular resolution of the observations (the full width at half maximum, FWHM, of the point spread function, PSF, assumed to be circular). We see no evidence of potential contamination from cirrus emission in the \Herschel/PACS $70\um$ image, confirming the circumstellar nature of the excess emission. Nearby background sources to the southeast and north become visible at $100\um$ and $160\um$, but these can be clearly differentiated from the extended emission of the circumstellar disk in the PACS images presented in Fig.~\ref{fig_herschelobs}. In the SPIRE sub-mm images, measurement of the disk extent and source flux is complicated due to the decreasing source flux and the blending of the background sources with the disk due to the larger telescope beam size at those wavelengths.

Scan map observations of HIP\,17439 were taken with both PACS 70/160 and 100/160 channel combinations. The observations were set up following the recommended parameters laid out in the scan map release note\footnote{see PICC-ME-TN-036, Section~1 for details.}, each scan map consisting of 10 legs of 3$'$, with a $4\arcsec$ separation between legs, scanning at the medium slew speed ($20\arcsec$ per second). Each target was observed at two array orientation angles ($70\degr$ and $110\degr$) to improve noise suppression and assist in the removal of low frequency (1/$f$) noise, instrumental artifacts and glitches from the images by mosaicking. A SPIRE observation in small map mode\footnote{see SPIRE observer's manual, Section~3.2.2 for details} was also taken of HIP\,17439, producing a fully sampled map covering a region $4'$ wide around the target. The \Herschell observations are summarized in Table~\ref{obs_log}.

\begin{table}
\caption{Summary of \Herschell observations of HIP\,17439.}             
\label{obs_log}      
\centering          
\begin{tabular*}{\linewidth}{l@{\extracolsep{\fill}}cccc}
\toprule
Inst. & Obs. IDs & OD\tablefootmark{a} & Wavelengths & OT\tablefootmark{b} \\
 & 13422... & & [$\mu$m] & [s] \\
\midrule
PACS  & ...22499/500 & 758 & 70/160      & 540   \\
PACS  & ...22501/502 & 758 & 100/160     & 1440  \\
SPIRE & ...14553     & 648 & 250/350/500 & 84    \\
\bottomrule
\end{tabular*}
\tablefoot{
\tablefoottext{a}{Operational day,}
\tablefoottext{b}{Total on-source integration time}
}
\end{table}

All data reduction is carried out in HIPE \citep[Herschel Interactive Processing Environment, ][]{ott10}, using calibration version 45 for the data reduction and user release version 10 for the PACS data and calibration version 10 for the SPIRE data. The individual PACS scans are processed with a high-pass filter to remove 1/$f$ noise, using high-pass filter widths of 15 frames at $70\um$, 20 frames at $100\um$ and 25 frames at $160\um$, suppressing structures larger than $62\arcsec$, $82\arcsec$ and $102\arcsec$ in the final images. For the filtering process, regions of the map where the pixel brightness exceeds a threshold defined as twice the standard deviation of the non-zero flux elements in the map are masked from the high-pass filter. De-glitching is carried out using the second level spatial deglitching task. The maps are created using the HIPE photProject task. The PACS scan pairs at $70\um$ and $100\um$, and the four $160\um$ scans (two each from PACS 70/160 and 100/160 observations) are mosaicked to produce the final images for each band. Final image scales are $1\arcsec$ per pixel at $70\um$ and $100\um$ and $2\arcsec$ per pixel at $160\um$, compared to native instrument pixel sizes of $3.2\arcsec$ at $70\um$ and $100\um$, and $6.4\arcsec$ at $160\um$, respectively. For the SPIRE observations, the small maps are created using the standard pipeline routine in HIPE and the na\"ive mapper option. Image scales of $6\arcsec$, $10\arcsec$ and $14\arcsec$ per pixel are used at $250\um$, $350\um$ and $500\um$. The sky background and rms uncertainty is estimated from the mean and standard deviation of the total fluxes in 25 boxes of $7\times7$ pixels at $70\um$, $9\times9$ pixels $100\um$ and $7\times7$ pixels at $160\um$ placed randomly around the center of the image. The box sizes are chosen to match the size of a circular aperture with maximum signal-to-noise at each wavelength which was determined from an analysis adopted to point-like sources. Since our source is extended and the aperture size is much larger than that the boxes are tailored for, we scale the measured per pixel noise to reflect the larger aperture area used to measure the fluxes of HIP\,17439. Regions are rejected if part of the box covers a region brighter than the threshold level as defined for the high-pass filtering and are spatially constrained to lie between $30\arcsec$ and $60\arcsec$ from the source position to avoid it and the lower coverage at larger distances from the center. The two nearby background sources are both fitted with a scaled and rotated (to account for the not perfectly circular shape of the telescope PSF with three side lobes the orientation of which depends on the telescope position angle during observations, here angle to be used is $-103.4\degr$) observation of $\alpha$\,Boo as a point source model and subtracted from the images before flux measurement.

The source flux in each of the three PACS images is measured using a circular aperture of radius $20\arcsec$. The aperture radii are chosen to encompass the full extent of the disk at all three wavelengths. The measured PACS fluxes are corrected for the aperture radius following PICC-ME-TN-037 (their Table~15). Color correction is applied as a combination of two temperature components (the stellar photosphere and the dust) using an interpolated value of the appropriate correction from Table~1 in PICC-ME-TN-038 after fitting a black body to the dust thermal emission to gauge its temperature. Corrections applied are 1.016, 1.034, and 1.074 at $70\um$, $100\um$, and $160\um$, respectively, for the star (the contribution of the photosphere at SPIRE wavelengths is negligible), and 0.982, 0.985, 1.010, 0.9417, 0.9498, and 0.9395 at $70\um$, $100\um$, $160\um$, $250\um$, $350\um$, and $500\um$, respectively, for the disk. Raw values of the fluxes are 73.3\,mJy, 90.0\,mJy, and 92.8\,mJy at $70\um$, $100\um$, and $160\,um$, respectively, and 53.0\,mJy, 32.2\,mJy, and 12.7\,mJy at $250\um$, $350\um$, and $500\um$, respectively. The PACS values are consistent with, but different from the ones in \citet{eir13}, because they have been derived with another version of the pipeline (differces in calibration version, pointing reconstruction and mosaicking). While the version used in the present work is newer, the values in \citet{eir13} are consistently derived for all targets and thus to be preferred when studying the whole sample instead of modeling a single object. Using the values in \citet{eir13} for the fitting presented in the present work would not alter the results significantly, but would slightly increase their significance.

At SPIRE wavelengths, the source is pointlike and the fluxes are measured in apertures of radii $22\arcsec$, $30\arcsec$, and $42\arcsec$ at $250\um$, $350\um$, and $500\um$, with the sky background estimated from an annulus of $60\arcsec-90\arcsec$ following the SPIRE aperture photometry recipe\footnote{SPIRE observer's manual, Section 5.2}. A summary of the optical and infrared photometry used in the SED fitting (including the new \Spitzerr and \Herschell fluxes) is presented in Table~\ref{phot}. In the SPIRE images the background sources are modeled as Gaussians with the appropriate beam FWHM for the wavelength of observation and are also subtracted before flux measurement.

Uncertainties on the photometry have been computed as the quadratic sum of sky noise and calibration uncertainties (5\% for PACS and 15\% for SPIRE, see \citealt{eir13} for details).

A full list of all photometric data of the HIP\,17439 system considered is given in Table~\ref{phot}.

\begin{table}
\caption{Optical and infrared photometry of HIP\,17439.}
\label{phot}
\begin{tabular*}{\linewidth}{l@{\extracolsep{\fill}}llcc}
\toprule
\multicolumn{1}{c}{$\lambda$} &\multicolumn{1}{c}{$F_{\rm obs}$}& \multicolumn{1}{c}{$F_{\rm pred}$} & Inst./ & Ref.\\
\multicolumn{1}{c}{$[\mu{\rm m}]$} & \multicolumn{1}{c}{[mJy]} &  \multicolumn{1}{c}{[mJy]} & source & \\
\midrule
\phantom{00}0.349 & $\phantom{00}682.9\pm\phantom{0}12.6$ & \phantom{00}\dots & a & 1    \\                     
\phantom{00}0.411 & $\phantom{0}2119.3\pm\phantom{0}39.0$           & \phantom{00}\dots & a & 1    \\
\phantom{00}0.440 & $\phantom{0}3057.8\pm\phantom{0}56.3$           & \phantom{00}\dots & b & 2    \\
\phantom{00}0.466 & $\phantom{0}4278.0\pm\phantom{0}78.8$           & \phantom{00}\dots & a & 1    \\
\phantom{00}0.546 & $\phantom{0}5924.3\pm109.1$                     & \phantom{00}\dots & a & 1    \\
\phantom{00}0.550 & $\phantom{0}5822.4\pm107.3$                     & \phantom{00}\dots & b & 2    \\
\phantom{00}0.790 & $\phantom{0}9430.6\pm173.7$                     & \phantom{00}\dots & b & 2    \\
\phantom{00}1.235 & $10416.0\pm182.3$                     & \phantom{00}\dots & c & 3    \\                      
\phantom{00}1.662 & $\phantom{0}9442.8\pm139.2$                     & \phantom{00}\dots & c & 3    \\                    
\phantom{00}2.159 & $\phantom{0}7085.9\pm104.4$                     & \phantom{00}\dots & c & 3    \\
\phantom{00}3.4   & $\phantom{0}3372.0\pm242.5$                     & \phantom{00}\dots & d & 4    \\ 
\phantom{00}9.0   & $\phantom{00}501.4\pm\phantom{00}7.9$ & 503.1           & e & 5      \\
\phantom{0}11.5   & $\phantom{00}314.4\pm\phantom{00}5.2$ & 309.5           & d & 4      \\
\phantom{0}12.0   & $\phantom{00}299.2\pm\phantom{0}17.9$ & 287.8           & f & 6      \\
\phantom{0}17.0   & $\phantom{00}154.7\pm\phantom{00}9.3$ & 145.2           & g & 7      \\
\phantom{0}22.1   & $\phantom{000}90.4\pm\phantom{00}2.0$ & \phantom{0}86.4 & d & 4      \\
\phantom{0}24.0   & $\phantom{000}67.8\pm\phantom{00}1.3$ & \phantom{0}73.3 & g & 7      \\
\phantom{0}25.0   & $\phantom{000}84.3\pm\phantom{0}13.5$ & \phantom{0}67.6 & f & 6      \\
\phantom{0}32.0   & $\phantom{000}53.1\pm\phantom{00}3.2$ & \phantom{0}41.3 & g & 7      \\
\phantom{0}60.0   & $\phantom{00}103.0\pm\phantom{0}24.7$ & \phantom{0}11.7 & f & 6      \\
\phantom{0}70.0   & $\phantom{000}99.1\pm\phantom{00}8.4$ & \phantom{00}8.6 & g & 7      \\
\phantom{0}70.0   & $\phantom{000}74.5\pm\phantom{00}3.8$ & \phantom{00}8.6 & h & 7      \\
100.0             & $\phantom{000}91.3\pm\phantom{00}4.7$ & \phantom{00}4.2 & h & 7      \\
160.0             & $\phantom{000}91.9\pm\phantom{00}4.9$ & \phantom{00}1.6 & h & 7      \\
250.0             & $\phantom{000}56.3\pm\phantom{0}10.5$ & \phantom{00}0.7 & i & 7      \\
350.0             & $\phantom{000}33.9\pm\phantom{00}8.9$ & \phantom{00}0.3 & i & 7      \\
500.0             & $\phantom{00}(13.5\pm\phantom{00}7.2$) & \phantom{00}0.2 & i & 7      \\
\bottomrule
\end{tabular*}
\tablefoot{Instruments/sources are: (a)~Stromgren~$ubvy$, (b)~from the \textit{HIPPARCOS} catalog: Johnson B, V, and Cousins I, (c)~2MASS~$JHK_\textrm{s}$, (d)~\textit{WISE}, (e)~\textit{AKARI}, (f)~\textit{IRAS}, (g)~\Spitzer, (h)~\Herschel/PACS, (i)~\Herschel/SPIRE. Uncertainties are 1-$\sigma$ and include contributions from both calibration and rms sky noise. Uncertainties on the predicted stellar photospheric fluxes are $1.5\%$. The SPIRE flux at $500\um$ can be considered an upper limit of 21.6\,mJy ($3\sigma$), while for the modeling we included it as the actual measurement on the marginal detection of the source and $1\sigma$ uncertainties for consistency with the other data.}
\tablebib{(1)~\citet{hau98}; (2)~\citet{per97}; (3)~\citet{skr06}; (4)~\citet{cut12}; (5)~\citet{ish10}; (6)~\citet{mos90}; (7)~This work.}
\end{table}

\subsection{Observational results}

We find the far-infrared emission to be associated with the star and interpret it as a circumstellar debris disk. The emission is extended compared to the PACS beam. From the SED we find that the emission is dominated by cold dust that must be located not too close to the star (significant emission starts between $24\um$ and $32\um$, black body grains peaking at these wavelengths would be located at $1.35\AU$ and $1.56\AU$ from the star, respectively, which can be considered as a lower limit of the inner edge of the disk). We find no evidence for significant amounts of warm dust at few au from the star or less at the sensitivity of the available photometry. From a black body fit to the disk's SED we find a temperature of $45.8\,\textrm{K}$ and fractional luminosity $L_\textrm{disk}/L_\textrm{star} = 10^{-4}$, which represent the first direct measurement of these values for the HIP\,17439 disk (note the lower limit derived by \citet{koe10} based on the \Spitzerr $70\um$ flux only which is in good agreement with our value). In the following, we discuss the constraints that can already be put on the disk structure from the spatially resolved images. We will present deconvolved images of the debris disk as well as radial profiles that will be used later for model fitting.

\subsubsection{Disk geometry and deconvolved images}
\label{sect_geom}

\begin{figure*}
\centering
\includegraphics[angle=0,width=1\linewidth]{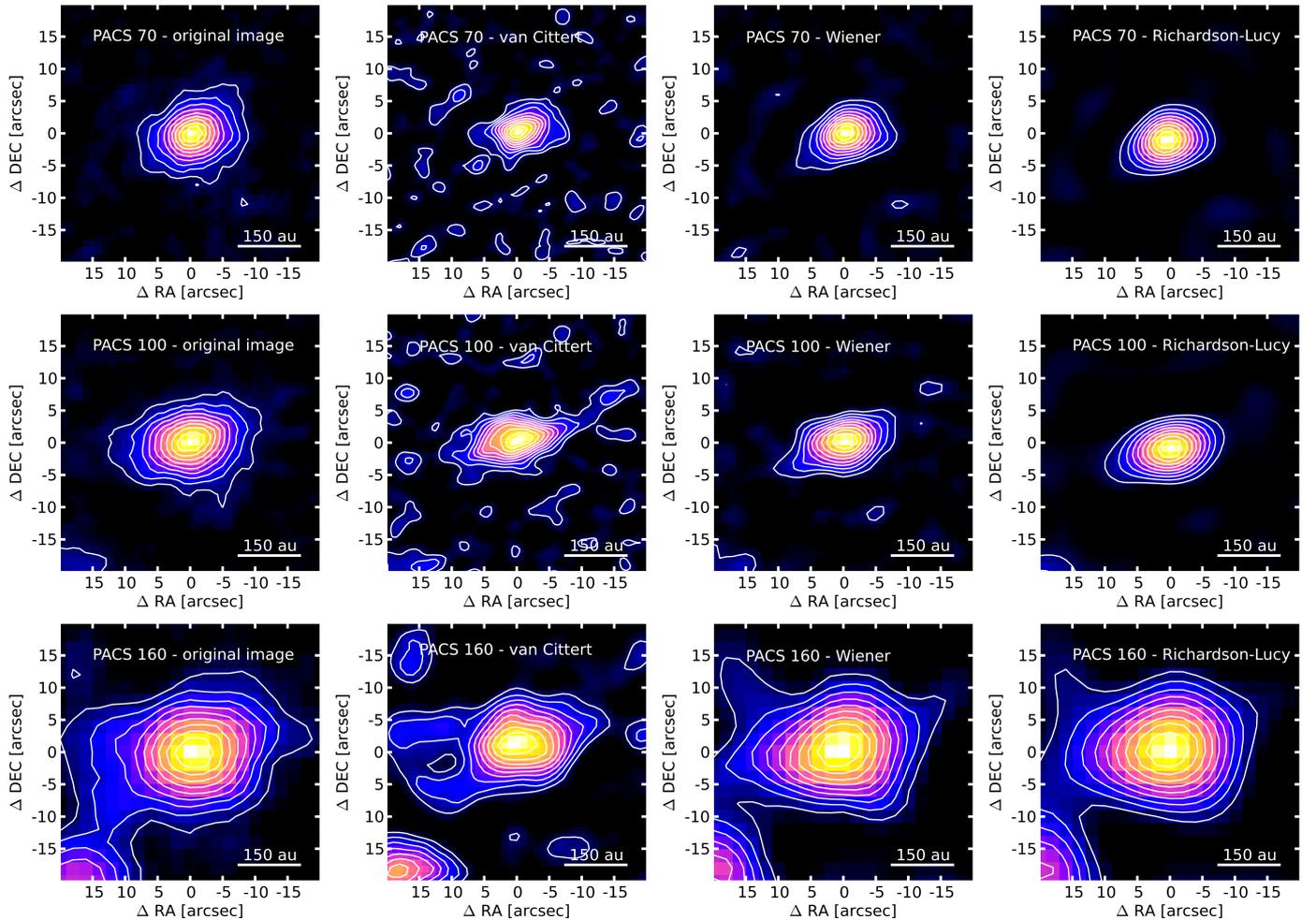}
\caption{Original images and deconvolved images at the three \Herschel/PACS bands. Image type, wavelength and deconvolution method are noted in the upper left corner of each panel, respectively. The images are oriented North up, East left. Contours denote positive surface brightness values and are drawn in steps of 10\% of the peak surface brightness. The color scale is linear and ranges from zero to the peak surface brightness.}
\label{fig_decon}
\end{figure*}

To derive a first estimate of the radial extent of the disk, its position angle, and its inclination, we fit a two dimensional Gaussian to the disk images at $70\um$, $100\um$, and $160\um$. In the images at SPIRE wavelengths, the disk is only marginally resolved (or not resolved at all). Thus, we do not consider these images. From the FWHM $A_\textrm{image}$ of the disk image at each wavelength measured along the major axis, we estimate the disk extent $A_\textrm{disk}$ (the extent of the structure dominating the emission at each wavelength) through deconvolution using the following formula (convolving a Gaussian with another Gaussian):
\begin{equation}
 \label{equ_decon}
 A_\textrm{image}^2 = A_\textrm{beam}^2 + A_\textrm{disk}^2~~~,
\end{equation}
where $A_\textrm{beam}$ is the FWHM of the beam. While $A_\textrm{disk}$ measured along the major axis of an image is representative of the diameter of the structure dominating the emission at this wavelength, the ratio between the deconvolved FWHM derived along the major axis, $A_\textrm{disk}$, and the minor axis, $B_\textrm{disk}$ (derived from the FWHM of the image along the minor axis, $B_\textrm{image}$), gives an estimate of the disk inclination $\theta$ from face-on (assuming a geometrically thin, circular disk):
\begin{equation}
 \cos(\theta) = \frac{B_\textrm{disk}}{A_\textrm{disk}}~~~.
\end{equation}
Note that this inclination has to be considered as a lower limit due to the limited angular resolution of the observations. The results of this analysis are listed in Table~\ref{tab_geom}.

\begin{table}
 \caption{Disk geometry measured from the PACS images}
 \label{tab_geom}
 \begin{center}
  \begin{tabular*}{1.0\linewidth}{l@{\extracolsep{\fill}}ccc}
   \toprule
Parameter                        &      $70\um$    &      $100\um$   &     $160\um$    \\
   \midrule
$A_\textrm{image}$ [$\arcsec$]        &   $9.9\pm0.2$   &  $13.4\pm0.1$   & $19.4\pm0.3$    \\
$B_\textrm{image}$ [$\arcsec$]        &   $7.7\pm0.1$   &  $8.7\pm0.1$    & $13.8\pm0.2$    \\
$A_\textrm{beam}$ [$\arcsec$]         &   $5.6$         &  $6.8$          & $11.3$          \\
$A_\textrm{disk}$ [$\arcsec$]         &   $8.2\pm0.3$   &  $11.5\pm0.1$   & $15.8\pm0.4$    \\
$A_\textrm{disk}$ [au]           &   $131.2\pm3.9$ &  $184.0\pm1.6$  & $252.8\pm6.0$   \\
$B_\textrm{disk}$ [$\arcsec$]         &   $5.3\pm0.2$   &  $5.4\pm0.2$    & $7.9\pm0.4$     \\
$\theta$ [deg]                   &    $\geq49.7$        &  $\geq62.0$     & $\geq60.0$          \\
$PA$ [deg]                       & $111.1\pm2.3$   & $103.2\pm0.9$   & $105.1\pm1.5$   \\
   \bottomrule
  \end{tabular*}
 \end{center}
 \tablefoot{The uncertainties on the disk extent represent the formal uncertainties of the measurements only and do not include any uncertainties originating from the method itself.}
\end{table}

Trying to resolve the dust depletion expected in the inner regions of the disk and in order to search for potential disk structures at the highest possible angular resolution, we perform image deconvolution on the PACS images. This deconvolution is a two-step process. Firstly, the stellar photosphere contribution is removed from each image by subtraction of a PSF rotated to the same position angle as the observations with a total flux scaled to the predicted photospheric level in that image centered on the stellar position determined from isophotal fitting of the 70~$\mu$m image. Secondly, after star subtraction, the image is deconvolved using an implementation of the van Cittert algorithm and the IRAF \textit{stsdas} packages for modified Wiener and Richardson-Lucy deconvolution to check the suitability of the individual methods and the repeatability of any structure observed in the deconvolved images. The observation of $\alpha$\,Boo adopted as a model PSF for the deconvolution process is assumed to be representative of a point source. However, analysis by \citet{ken12} have identified variations at the 10\% level in the extent of the PSF FWHM at $70\um$ for a variety of point source calibrators depending on the date and mode of observation (a smaller variation of 2-4\% is seen at $100\um$). Variation at this level does not negate the interpretation of this target as an extended source, but should be kept in mind to avoid over interpretation of the extended structure revealed through the deconvolution process or of the results of the model fitting presented later, as the model images are concolved with the same PSF images in order to produce simulated observations. A comparison of the extent of the disk at different wavelengths in the PACS images is still valid, as the data have been obtained at the same day in the same observing mode.

A comparison of the three deconvolution methods used in this work is provided in Fig.~\ref{fig_decon}. No significant structure or asymmetry is visible in these deconvolved images. Furthermore, despite the increased angular resolution, no dip or plateau of the surface brightness in the inner regions that would be indicative of an inner hole is visible. Thus, we conclude that the diameter of any depleted inner region of the disk is not significantly larger than the resolution reached after deconvolution (typically about a factor of two improvement over original resolution, i.e., approx.\ $50\AU$ at $70\um$).

Furthermore, it becomes obvious that the extent of the disk measured at different wavelengths is significantly different according to both Equ.~\ref{equ_decon} and image deconvolution (a factor of two from $70\um$ to $160\um$ according to Equ.~\ref{equ_decon}). The possibility that this is caused by the lower angular resolution at longer wavelengths and imperfect deconvolution cannot be ignored. In contrast, in previous, similar data of HD\,207129 \citep{loe12}, we found for a disk with a confined, ring-like shape a good agreement between the extent measured at different wavelengths. Thus, we conclude that our present findings are indicative of a radially very extended disk with a shallow surface density profile (colder dust further away from the star emits more efficiently at longer wavelengths) or of a disk composed of multiple rings at significantly different radial distance from the star, and with flux ratios that vary significantly with wavelength in the \Herschel/PACS spectral range. We will further discuss these scenarios in the modeling section.

\subsubsection{Radial profiles}

\begin{figure*}
\centering
\includegraphics[angle=0,width=0.92\linewidth]{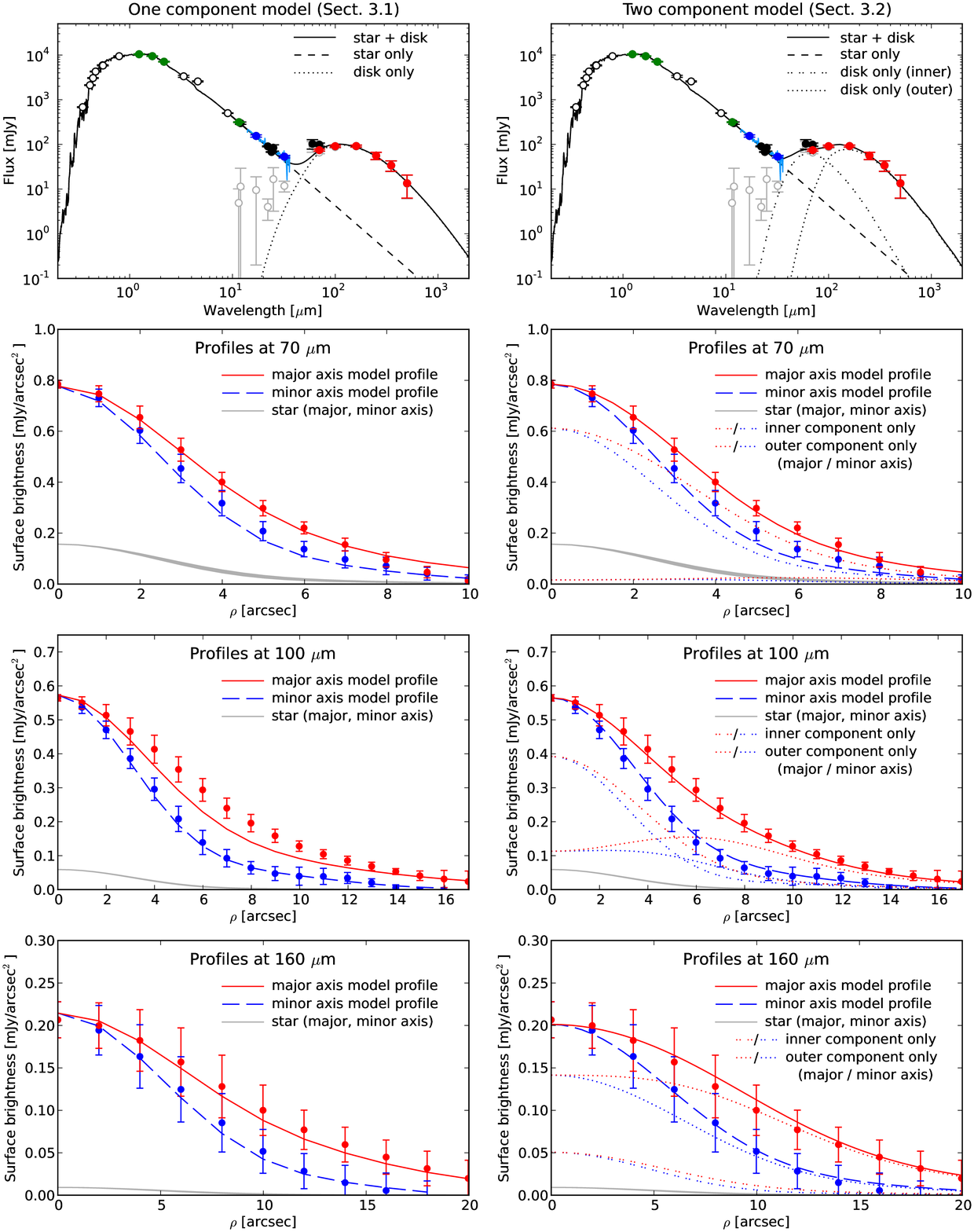}
\caption{Observed SED and radial profiles plotted along with simulated data from our one-component (\emph{left column}) and two-component best-fit models (\emph{right column}). \emph{In the SED plots}, red data points are our new \Herschell data, the light blue line represents the available \Spitzer/IRS spectrum, blue points represent synthetic photometry extracted from this spectrum in order to consider them for the model fitting, gray points illustrate the photosphere subtracted photometry longward of $10\um$, and remaining points are ancillary data from the literature. Green points represent the data used for the normalization of the stellar spectrum. Data longward of $10\um$ have been considered for the disk model fitting. \emph{In the radial profile plots}, the thick black line represents the stellar contribution (an observation of the PSF reference star $\alpha$\,Boo scaled to the stellar flux of HIP\,17439), with its width representing the difference between major axis and minor axis. The remaining lines represent the model data and points with uncertainties represent the observed profiles. The major axis is plotted in red, the minor axis in blue. \emph{In the right column} showing the results for the two-component fit, dotted lines illustrate the contribution of the two-components (inner/warm and outer/cold).}
\label{fig_fits_onetwo}
\end{figure*}

Radial profiles are useful to visualize extended emission. Furthermore, we use radial profiles in our simultaneous multi-wavelength fitting of the data presented later in this paper. We produce radial profiles of the source for each PACS wavelength in the following manner: A cutout of the mosaic image at each wavelength is centered on the stellar position and the disc position angle is determined by fitting a 2D Gaussian to the source brightness profile at $100\um$ (same procedure as in the previous section). Each cutout is then rotated such that the disk major axis lies along the rotated image's $x$-axis. The cutout is then interpolated using the IDL INTERPOLATE routine to a grid with ten times the density of points, equivalent to spacings of $0.1\arcsec$ per element at $70\um$ and $100\um$ and $0.2\arcsec$ per element at $160\um$ (rebinning by a factor of 10 for each wavelength). The (sub-) pixel values of two regions of this interpolated image on opposite sides of the center, covering 11$\times$11 pixels on a side are averaged at distance intervals equivalent to $1\arcsec$ from the source center along the image $x$-axis at $70\um$ and $100\um$ and at intervals of $2\arcsec$ at $160\um$ to derive the values of the radial profile. The uncertainty is taken by combining the sky noise (determined from aperture photometry) and difference in the mean of the two sub-regions in quadrature. The same method is applied along the y-axis to generate the radial profile of the source minor axis. The model PSF images are processed in the same manner after rotation to the same position angle as the observations to provide a point source radial profile for comparison. Due to the decreasing source flux, contamination from background sources, larger rms background uncertainties, broader instrument PSF and lower spatial sampling of the source radial profile we do not consider the images at SPIRE wavelengths (being consistent with the point source model). The resulting radial profiles are shown in Fig.~\ref{fig_fits_onetwo}.

\section{Analytical multi-wavelength modeling of the disk}
\label{sect_mod}

\begin{table*}
 \caption{Parameter space explored and best-fit values for the one-component model (\emph{top}), and the two-component model (\emph{bottom}).}
 \label{tab_fits_general}
 \begin{center}
  \begin{tabular*}{\linewidth}{l@{\extracolsep{\fill}}ccccc}
   \toprule
Parameter   &  Range explored & \# values & Distribution & Best-fit [$3\sigma$ confidence levels] & Component \\
   \midrule
\multicolumn{6}{c}{\underline{\textit{One-component model (Sect.~\ref{sect_one_comp})}}} \\
   \midrule
$R_\textrm{in}$ [au]     & $5.0~...~100.0$     & $400$     & temp     & $8.3$ [$7.5~...~13.9$]    & N/A \\
$R_\textrm{out}$ [au]    & $15.0~...~500.0$    & $246$     & temp     & $394.0$ [$126.6~...~500^*$]  & N/A \\
$\alpha$                 & ${-5.0}~...~2.5$    & $76$      & lin      & ${-0.1}$ [${-1.6}~...~0.9$]    & N/A \\
   \midrule
$a_\textrm{min}$ [$\um$] & $0.2~...~20.0$      & $449$     & log      & $8.1$ [$6.2~...~10.7$]     & N/A \\
$a_\textrm{max}$ [$\um$] & $2000.0$            & $1$       & fixed    & \dots                      & N/A \\
$\gamma$                 & ${-2.0}~...~{-5.0}$ & $31$      & lin      & ${-4.0}$ [${-3.0}^*~...~{-4.9}^*$] & N/A \\
   \midrule
$\theta$ [deg]           & $0.0~...~90.0$      & $51$      & cos      & $63.9$ [$17.8~...~82.0$]   & N/A \\
$M_\textrm{1mm}$ [$M_\oplus$] & \dots         & \dots     & cont     & $1.3\times10^{-2}$         & N/A \\
   \midrule
$\tau_\bot(R_\textrm{in})$      & \dots & \dots & \dots & $5.5\times10^{-5}$ & N/A \\
$\tau_\bot(R_\textrm{out})$     & \dots & \dots & \dots & $3.7\times10^{-5}$ & N/A \\
$t_\textrm{coll}(R_\textrm{in}) $ [yr]  & \dots & \dots & \dots & $8.1\times10^{4}$  & N/A \\
$t_\textrm{coll}(R_\textrm{out})$ [yr]  & \dots & \dots & \dots & $3.9\times10^{8}$  & N/A \\
$t_\textrm{PR}(R_\textrm{in})   $ [yr]  & \dots & \dots & \dots & $7.5\times10^{4}$  & N/A \\
$t_\textrm{PR}(R_\textrm{out})  $ [yr]  & \dots & \dots & \dots & $1.7\times10^{8}$  & N/A \\
   \midrule
$\chi^2_\textrm{red}$    & \dots               & \dots     & \dots    & 2.35                       & N/A \\
   \midrule
\multicolumn{6}{c}{\underline{\textit{Two-component model (Sect.~\ref{sect_two_comp})}}} \\
   \midrule
$R_\textrm{in,1}$ [au]     & $1.0~...~150.0$     & $50$      & log      & $29.2$ [$1.7~...~35.8$]    & inner \\
$R_\textrm{out,1}$ [au]    & $500.0$             & $1$       & fixed    & \dots                      & inner \\
$\alpha_\textrm{1}$        & ${-5.0}~...~2.0$    & $36$      & lin      & ${-4.0}$ [${-0.4}~...~{-5.0}^*$] & inner \\
$R_\textrm{in,2}$ [au]     & $10.0~...~200.0$    & $20$      & log      & $90.9$ [$16.0~...~170.8$] & outer \\
$R_\textrm{out,2}$ [au]    & $500.0$             & $1$       & fixed    & \dots                      & outer \\
$\alpha_\textrm{2}$        & ${-5.0}~...~2.0$    & $36$      & lin      & ${-1.6}$ [${1.0}~...~{-5.0}^*$] & outer \\
   \midrule
$a_\textrm{min,1}$ [$\um$] & $0.05~...~43.5$     & $55$      & log      & $5.2$ [$3.5~...~16.0$]     & inner \\
$a_\textrm{max,1}$ [$\um$] & $2000.0$            & $1$       & fixed    & \dots                      & inner \\
$\gamma_\textrm{1}$        & ${-2.5}~...~{-5.5}$ & $16$      & lin      & ${-5.5}^*$ [${-3.7}~...~{-5.5}^*$] & inner \\
$a_\textrm{min,2}$ [$\um$] & $0.05~...~43.5$     & $55$      & log      & $12.4$ [$0.05^*~...~29.9$] & outer \\
$a_\textrm{max,2}$ [$\um$] & $2000.0$            & $1$       & fixed    & \dots                      & outer \\
$\gamma_\textrm{2}$        & ${-2.5}~...~{-5.5}$ & $16$      & lin      & ${-4.3}$ [${-3.0}~...~{-5.5}^*$] & outer \\
   \midrule
$\theta$ [deg]             & $50.0~...~70.0$     & $5$       & lin      & $60.0$ [$50.0^*~...~70.0^*$]   & both \\
$V_\textrm{ice}/\left(V_\textrm{ice}+V_\textrm{si}\right)$ [\%]& $0.0~...~50.0$ & $2$ & lin & $0.0$ [$0.0~...~50.0$] & both \\
   \midrule
$M_\textrm{1mm,1}$ [$M_\oplus$] & \dots         & \dots     & cont     & $2.1\times10^{-4}$         & inner \\
$M_\textrm{1mm,2}$ [$M_\oplus$] & \dots         & \dots     & cont     & $1.1\times10^{-2}$         & outer \\
   \midrule
$\tau_\bot(R_\textrm{in,1})$     & \dots & \dots & \dots & $9.3\times10^{-4}$ & inner \\
$\tau_\bot(R_\textrm{in,2})$     & \dots & \dots & \dots & $2.6\times10^{-4}$ & outer \\
$t_\textrm{coll}(R_\textrm{in,1})$ [yr]   & \dots & \dots & \dots & $6.6\times10^{4}$  & inner \\
$t_\textrm{coll}(R_\textrm{in,2})$ [yr]   & \dots & \dots & \dots & $1.3\times10^{6}$  & outer \\
$t_\textrm{PR}(R_\textrm{in,1})  $ [yr]   & \dots & \dots & \dots & $9.4\times10^{5}$  & inner \\
$t_\textrm{PR}(R_\textrm{in,2})  $ [yr]   & \dots & \dots & \dots & $9.1\times10^{6}$  & outer \\
   \midrule
$\chi^2_\textrm{red}$      & \dots & \dots & \dots & 1.67               & both  \\
   \bottomrule
  \end{tabular*}
 \end{center}
 \tablefoot{The distributions of the values considered in the parameter space are: \emph{temp} -- by equal steps in temperature of the grain size with the steepest radial temperature gradient, \emph{lin} -- linear, \emph{log} -- linear in the logarithm of the parameter, \emph{fixed} -- fixed value (no distribution at all), \emph{cos} -- linear in the cosine of the parameter, \emph{cont} -- continuous (scaling of the disk mass to minimize the $\chi^2$ for given values of all other parameters). Values marked with an asterisk are (very close to) the boundaries of the parameter space explored and cannot be considered as reliable (in case of best-fit parameters real values may lie outside the parameter space explored, in case of confidence levels uncertainties are probably larger here). The values for $\tau_\bot$ (vertical geometrical optical depth), and $t_\textrm{coll}$ (collisional lifetime of the dust following \citealt{bac93}), and $t_\textrm{PR}$ (Poynting-Robertson life time of dust grains, see Sect.~\ref{sect_one_comp} for details) are computed for the best-fit models at representative positions in the disk.}
\end{table*}

In the previous section, we found indications that the disk is either radially very extended (from $<$~$65\AU$ to $>$~$125\AU$, see values for $A_\textrm{disk}$ in Table~\ref{tab_geom}), or is composed of multiple belts of dust. A final conclusion was, however, not possible. To investigate the two scenarios and to further constrain the dust spatial distribution and composition, we now perform detailed analytical\footnote{Here, the term `analytical' refers to fitting of the observational data to spatial and size distributions of dust, assumed to be power laws. It is used to distinguish this kind of modeling from collisional and dynamical simulations that address physical processes of the dust production and evolution and deal with dust distributions that are more complex than power laws.} model fitting to the SED and the radial profiles of the system simultaneously.

\subsection{Single-component model with variable radial width}
\label{sect_one_comp}

The standard approach for modeling spatially resolved data of debris disks used in the context of \Herschel/DUNES employs a disk model composed of a single disk component \citep{loe12, ert12diss}. The one-component model we consider here is described by two power-laws. The spatial distribution of the dust is described by a power-law radial surface density distribution (exponent $\alpha$) with inner radius $R_\textrm{in}$ and outer radius $R_\textrm{out}$. The differential size distribution of the dust grains is described by a power-law (exponent $\gamma$) with lower grain size $a_\textrm{min}$ and upper grain size $a_\textrm{max}$. In the present work, $a_\textrm{max}$ has been fixed to a reasonably large value of 2\,mm, so that the effect of this parameter is negligible for reasonably steep grain size distributions ($\gamma < -3.0$). Although larger values of $\gamma$ are considered in the parameter space explored, our best-fit values lie well below $-3.0$. Absorption coefficients of the dust grains (assumed to be spherical and compact) are computed employing Mie theory assuming the dust to be composed of astronomical silicate \citep{dra03}. Composition and size distribution of the dust grains are thereby assumed to be the same throughout the whole disk. The dust mass $M_\textrm{1mm}$ in grains up to 1\,mm (different from the value of $a_\textrm{max}$ because 1\,mm is a common value used in the literature that permits easier comparison with other studies) and the disk inclination $\theta$ from face-on orientation completes the suit of model parameters. For the fitting we consider a total of 98 data points (SED and radial profile points as illustrated in Fig.~\ref{fig_fits_onetwo}).

To fit this model to the data, we use the tool \texttt{SAnD} described in detail by \citet{ert12a} and \citet{ert12diss}, and applied to spatially resolved data in \citet{loe12} and \citet{ert12diss}. It uses a simulated annealing approach to find the global best-fit parameters in a high dimensional parameter space. The parameter space explored in this approach is summarized in Table~\ref{tab_fits_general} along with the best-fit parameters found. The resulting best-fit SED and profiles are shown in the left column of Fig.~\ref{fig_fits_onetwo}.

Comparing the data observed and simulated from our one-component best-fit model, we find that most observations are well reproduced by the model. However, there is a significant deviation for the $100\um$ profile along the major axis. Along this axis, the modeled profile is significantly less extended than the observed one. Nonetheless, the overall reduced $\chi^2$ is reasonably low ($2.35$) and considering the whole data set, this deviation may or may not be considered as critical.

A more important indication that the model used is unable to reproduce the data in a reasonable way is the set of best-fit parameters derived. These parameters indicate a very broad disk ranging from a few au up to few hundreds of au from the star with a surface density that is approximately constant. The formation of such a disk would be hard to explain. A constant surface density would be expected from a transport dominated disk (spatial distribution of the dust due to Poynting-Robertson drag dominates over local production due to collisions). Table~\ref{tab_fits_general} lists the collisional life time $t_\textrm{coll}$ and Poynting-Robertson life time $t_\textrm{PR}$ for our model at the inner and outer radius of the disk. Poynting-Robertson life times are computed following \citet{gus94} assuming grains with $\beta = 0.5$ with $\beta$ being the ratio between radiation pressure force and gravitational force. Indeed, $t_\textrm{PR}$ and $t_\textrm{coll}$ of the dust in our best-fit are of the same order, suggesting that transport mechanisms may have a significant contribution to the dust dynamics.

Estimating lifetimes for $\beta = 0.5$ grains, we implicitly assumed the grains close to the blowout size to be the most abundant in the disk. However, the grains around HIP 17439 may not reach beta = 0.5, at least assuming spherical compact grains of standard composition (e.g. astronomical silicate or silicate-ice mixtures; e.g., \citealt{kir13}). As a result, even smaller, submicron-sized grains may be able to stay in bound orbits drifting inward to the star. \citet{rei11} have demonstrated how disks around low-luminosity stars, where the blowout limit does not exist, may become transport-dominated even at high levels of the optical depth and thus relatively short collisional lifetimes. This may become even more efficient, if additional processes such as stellar winds are operating in the system.

However, the transport-dominated scenario faces some difficulties. One of them is that, whereas transport should be the most efficient for submicron sizes, the smallest grains found by our fitting are a few microns in size. This discrepancy can potentially be mitigated in several ways, for instance by changing assumptions about the grains properties (chemical composition, porosity, etc.). Another, more serious difficulty is that in this scenario the majority of the dust would have to be produced through collisions of larger bodies close to the outer edge of the disk, suggesting a significant amount of planetesimals at few hundreds of au from the star ($\sim$$200\AU$ or larger considering the uncertainties on $R_\textrm{out}$), which at least remains questionable. While an even smaller $R_\textrm{out}$ would still be possible within the error bars, this needs to be compensated by changes in other parameters such as an outwards increasing surface density slope.

An alternative explanation to the transport dominated disk would be that the dust is produced locally throughout the whole disk with a local production rate that results in a constant surface density. To check whether this is feasible, we employ an analytic model of dust production in a steady-state collisional cascade, operating in a planetesimal disk \citep{loe08}. We assume that the planetesimal disk initially had a solid surface density of the standard Minimum Mass Solar Nebula (MMSN) model \citep{wei77, hay81}, $\Sigma = 1M_\oplus \textrm{au}^{-2} (r/\textrm{au})^{-3/2}$, and consider two possible mechanisms of the cascade activation. One is self-stirring, in which the cascade is assumed to start at a distance $r$ at the moment when the $1000\,\textrm{km}$-sized planetesimals form there \citep{ken08}. In another case the cascade is ignited by the stirring front from a planet residing inside the disk \citep{mus09}. The mass of the central star is set to $M_\star = 0.7\,M_\odot$. We assume that 10\% of the available MMSN mass in solids went into formation of planetesimals up to $100\,\textrm{km}$ in radius. The ``primordial'' slope of the differential mass distribution of planetesimals (i.e., that of the planetesimals that are sufficiently big to have collisional lifetimes longer than the time elapsed from the cascade ignition) is taken to be $q_\textrm{p} =1.87$ \citep[cf.][]{loe08}, which corresponds to a slope of the differential size distribution of $\gamma = -3.61$ assuming a constant bulk density of bodies of different size ($n(a)da \propto a^\gamma da \propto m^{2-3q}dm$). For the mean eccentricities and inclinations of dust-producing planetesimals after the ignition of the cascade, we adopt $e \sim 2I = 0.1$. The critical shattering energy $Q_D^\star$ is chosen as in \citet{ben99}. In the planetary stirring case, we assume a Jupiter-mass planet with a semi-major axis of $10\AU$ and an orbital eccentricity of $0.1$.

\begin{figure}
\centering
 \includegraphics[width=1.0\columnwidth, angle=0]{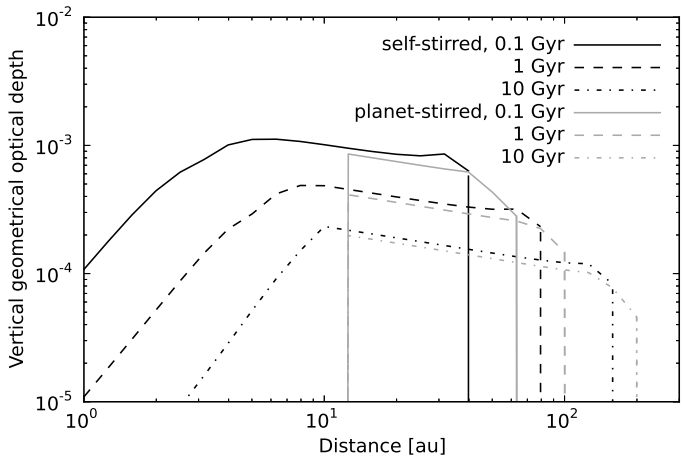}\\
 \caption{Radial profiles of the vertical geometrical optical depth of a steady-state debris disk stirred by embedded Pluto-size objects (self-stirred, black lines) and by a planet in the inner cavity of the disk (planet-stirred, gray lines), and at different time instants: after $0.1\Gyr$ (solid), $1\Gyr$ (dashed), and $10\Gyr$ (dot-dashed) after the beginning of the collisional evolution. The lines of the planet-stirred case have been scaled by a factor of 1.1 for better visibility.}
\label{fig_collisions}
\end{figure}
   
Typical results for one particular set of parameters, in the form of the radial profiles of the disk's vertical optical depth $\tau_\bot$, are presented in Fig.~\ref{fig_collisions}. The inner parts of the modeled disk ($\la 10\AU$) have $\tau_\bot$ increasing outward with a slope of $7/3$, as expected in the case where the collisional age of the system is longer than the collisional lifetime of the largest planetesimals \citep[see, e.g.,][]{ken10, wya12}. Farther out from the star, the profiles change to a nearly horizontal plateau, whose outer edge corresponds to the current position of the stirring front. Both the knee between the regions of increasing and constant $\tau_\bot$ and the outer edge of the disk move to larger distances at later times. Thus, these simulations seem to be able to yield broad disks with a nearly constant $\tau_\bot$. However, it proves very difficult to get a disk extending to more than $\sim$$100\AU$, as required to reproduce the observations of the HIP\,17439 disk. Varying the parameters of the model would not substantially change the results. Furthermore, as argued before against the scenario of a transport dominated disk, it is unlikely that sufficient mass in solids can be available at hundreds of au from the star. Even if it is, it would be difficult for Pluto-sized stirrers to form in the disk at those distances in $\sim$1\,Gyr. Moreover, in both planetary and self-stirring scenarios, at such large distances from the star only small bodies - which are in the strength regime and are harder to break - have collisional times shorter than the age of the system, and so only little dust is produced. This explains the second knee in Fig.4 where $\tau_\bot$ starts to drop (at $\sim$$120\AU$ in the 10\,Gyr curves). We note, however, that the current planetesimal formation models are very uncertain, being very sensitive, for instance, to the assumed initial sizes of the seed planetesimals \citep{keny10}. Besides, there exist alternative mechanisms to rapidly form sufficiently large stirrers at large distances from the star, based on concentration by instabilities followed by local clumping (e.g., \citealt{joh12}).

We summarize that we cannot rule out the broad ring model observationally due to the limited angular resolution and large uncertainties in the radial profiles. Furthermore, we cannot rule it out theoretically, since our modeling does not cover all possible scenarios, is limited by some assumptions, and not all parameters of the models have been fully explored (this is beyond the scope of this paper and we refer here to a subsequent paper by Sch\"uppler et al., in prep.). However, based on the combination of the facts that the fit of this model to the data is imperfect and that our modeling did not result in any plausible scenario that could explain such a broad, extended disk, we conclude that such a scenario is rather unlikely.

\subsection{Adding a second disk component}
\label{sect_two_comp}

The results from our first fitting approach demonstrate that a single power-law disk model with the same dust composition and grain size distribution over the whole disk is unable to reproduce all data simultaneously. This is because any such disk model that results in sufficiently extended emission at $100\um$ also results in significantly extended emission at $70\um$, which is not observed. Possible ways to improve over such a model would in general require an extension of it that can be considered as an additional disk component. In such a case, the outer component producing the extended emission at $100\um$ must have properties that result in very low emission at $70\um$. These might be a large lower grain size resulting in a low emissivity at this wavelength and/or in a low temperature, or a large distance from the star also resulting in a low temperature (the distance from the star is, however, constrained by the extent of the disk images).

To explore these possibilities, we set up a new model featuring two independent power-law disk components each of which is identical to the single-component model used before, with slightly different sets of free and fixed parameters, as described in the following and listed in Table~\ref{tab_fits_general}. These two components are weighted against each other and added up to the final model. Thus, for each component $i$ ($i \in \left[1,2\right]$) there is an inner disk radius $R_{\textrm{in}, i}$, an outer disk radius $R_{\textrm{out}, i}$, an exponent $\alpha_i$, a lower grain size $a_{\textrm{min}, i}$, an upper grain size $a_{\textrm{max}, i}$, an exponent $\gamma_i$, and a dust mass $M_{\textrm{1mm}, i}$ in grains up to 1\,mm in size. On the other hand, the inclination $\theta$ is assumed to be the same for both components. While this assumption might not be fulfilled, the $70\um$ emission (most constraining the orientation of any assumed inner component) is barely resolved and thus our assumption does not significantly affect the results. Furthermore, we now consider two possible compositions of the dust -- pure astronomical silicate \citep{dra03} and a 1:1 mixture of astronomical silicate and ice \citep{li98, loe12, ert12a}. The volume fraction $V_\textrm{ice}/\left(V_\textrm{ice}+V_\textrm{si}\right)$ of water ice in the silicate and ice dust grains is initially assumed to be the same for both components. In addition, we consider sublimation of the water ice in a way that the ice is replaced by vacuum if the grain temperature exceeds the sublimation temperature of the ice \citep{leb12}. Finally, we fix the outer radius of both components to $500\AU$. This allows us to further limit the number of free parameters to be fitted. The consequence in the model is that the surface density profile of both components is most probably forced to be decreasing outwards which results in two overlapping disk components that can be interpreted as two belts where the emission of each belt is dominated by its inner rim, respectively. The values of $\alpha_i$ are then a measure of the width of each belt, and if the slope of the surface density of the inner belt is steep enough, the two components can be considered as detached. This way we have a total of 12 parameters to be fitted, of which $M_{\textrm{1mm}, i}$ (technically, the total mass and the mass ratio of the two components) are determined for each set of the other 10 parameters using a downhill $\chi^2$ minimization.

In order to explore a 10-dimensional parameter space, we use a hybrid approach between a precomputed model grid and the simulated annealing approach of \texttt{SAnD}. We use the tool \texttt{GRaTer} \citep{aug99,leb12,loe12} to compute and store a grid of model data \textit{for each of the two components separately}. On these grids of precomputed model data, we perform model fitting using a simulated annealing approach analogous to that used by \texttt{SAnD}. We combine for each step of the simulated annealing random walk one si\-mulated data set from each of the two model grids and fit them to the data after proper scaling. Uncertainties on the best-fit parameters are estimated analogous to the approach used in \texttt{SAND}. In order to not confuse the two components in the error estimation, we ensure that $R_\textrm{in,1} < R_\textrm{in,2}$ for each step.

The effectively explored parameter space, the resulting best-fit parameters, and the uncertainties on them are listed in Table~\ref{tab_fits_general}. The SED and radial profiles simulated for this best-fit model are shown in the right panel of Fig.~\ref{fig_fits_onetwo}. It is important to note that given the high dimensional and extremely complex parameter space and the fact that we can only perform a limited number of runs in a reasonable time, we cannot claim anymore having found the global best-fit in the parameter space. However, we performed in total 10 independent runs with different limitations of the parameter space, all end up in the same sink of $\chi^2$ in the parameter space and do not find a deeper one.

We are able to significantly improve the fit to the $100\um$ major axis profile without significantly degrading the fit to the other data. The constraints on the disk parameters are in general weak in this approach due to the intrinsic degeneracies of the model and the low spatial resolution of the data. Furthermore, models similar to our one-component best-fit model are included in the parameter space of both components. In the previous section, we already found that this model results in a fit of limited, but still acceptable quality ($\chi^2 = 2.35$). Thus, one of the two components in the present approach alone can reproduce the data sufficiently within $3\sigma$ uncertainties of the model parameters, in which case the other component can assume nearly arbitrary values. This is a major source of the large $3\sigma$ uncertainties on the model parameters. The best-fit two-component model consists of two well detached components both of which are extended disks with inner radii of $\sim$$30\AU$ and $\sim$$90\AU$, respectively, and with outwards decreasing surface density profiles. Both components are collision dominated. The vertical optical depth is by a factor of $\sim$$3.5$ higher for the inner component compared to the outer one. Deviations from standard parameters expected from such fits are the rather steep grain size distributions and the large lower grain size in both components, in particular in the outer component. However, the uncertainties for all these parameters are large and nearly all parameters are consistent with standard values (micron or sub-micron sized for the smallest grains, a standard value of $-3.5$ for the slope of the differential size distribution) within $3\sigma$ fitting uncertainties.

The fact that the constraints on the parameters of the best-fit two-component model are weak raises the question whether a normal ring of debris at few tens of au and an additional halo of small, barely bound grains produced in this ring and drawn to large orbits by radiation pressure or stellar winds might reproduce the data as well. While the grains in such a halo should be the smallest grains present in the system, we note that our best-fit result of the two-component model suggests that the emission in the outer disk component is most likely dominated by larger grains than that of the inner component. While we cannot completely rule out that the outer disk is composed mostly of small grains due to the barely constrained outer disk parameters, we rate this scenario as very unlikely. Such small grains ($a \sim 1\um$) barely emit at wavelengths as long as $100\um$, but emit more efficiently at shorter wavelength. This is inconsistent with our observation that the outer disk becomes visible only at $100\um$ and longer, as long as the other model parameters do not assume very extreme values like a constant surface density, which would then be again inconsistent with a halo. Furthermore, given the luminosity of the star being at the border to not producing blow-out grains at all, it heavily depends on the parameters of the dust (e.g., composition, density, porosity, e.g., \citealt{kir13}) whether there are barely bound grains at all.

\section{Conclusions}
\label{sect_conc}

We have spatially resolved for the first time the debris disk around HIP\,17439. The disk morphology changes significantly from $70\um$ to $100\um$ suggesting a very extended disk or a disk composed of multiple components to be present. Our simultaneous multi-wavelength modeling of all available data of this disk including the radial profiles obtained from our resolved observations shows that a single-component model does reproduce most of the data in a reasonable way but fails in reproducing the profiles at $70\um$ and $100\um$ simultaneously, i.e., in reproducing the change in disk morphology between these two wavelengths. We furthermore have demonstrated with an analytical approach that the resulting best-fit model as well as most models allowed within $3\sigma$ of the uncertainties on the model parameters are physically unlikely. In contrast, we have demonstrated that a two-component disk model is able to reproduce all data simultaneously and results in a physically more plausible result. This best-fit model is composed of two rings of debris located at $\sim$$30\AU$ and at $\sim$$90\AU$ from the star. The \textit{dust mass} derived from our model puts the disk among the least massive debris disks spatially resolved so far. Depending on the disk geometry (which strongly affects the vertical optical depth, but could not be unambiguously constrained in the present work), transport mechanisms may play a significant role in the dust dynamics. This potentially makes the disk similar to that of HD\,207129 \citep{loe12}, and interesting to study dust dynamics and evolution in debris disks with low surface density. It is also particularly interesting as the disk does not have a particularly low fractional luminosity compared to other debris disks detected by Spitzer or Herschel \citep[e.g.,][]{bry06, bei06, hil08, bry09, eir13}. This means that in a significant fraction of known debris disks transport mechanisms might not be negligible. This is in contrast to the estimates by \citet{wya05}, who assumed debris disks to be narrow rings which might not be the case for HIP\,17439.

In the lack of higher spatial resolution, which would require a higher sensitivity to surface brightness than can be achieved with current instruments (including ALMA judged from our best-fit models, see predictions by \citealt{ert12b}, their Fig.~5, left) only spatially resolved \textit{multi-wavelength} observations and the detailed simultaneous modeling of all available data were able to reveal the multi-component (or possible extremely extended) structure of the disk. A more detailed modeling of the system's dust distribution considering self consistently the dynamical and collisional evolution of both the planetesimals and the dust in order to investigate whether the scenarios found are physically really plausible is beyond the scope of this paper. We postpone this to a later study (Ch.~Sch\"uppler et al., in prep.). This study will potentially reveal the conditions under which the scenarios found are physical and further constrain possible disk architectures, including the locations of dust-producing planetesimal belts analogous to previous studies for other objects \citep{rei11, loe12}.

In recent years, more an increasing number of debris disks have been modeled in detail and revealed to consist of several components, more or less similar to the architecture of our own Solar system (e.g., $\epsilon$\,Eri: \citealt{bac09}; HR\,8799: \citealt{su09}; $\eta$\,Crv: \citealt{mat10}; HD\,107146: \citealt{ert11}; Fomalhaut: \citealt{ack12}, \citealt{su13}; HD\,32297: \citealt{don13}; Vega: \citealt{su13}; $\gamma$\,Dor: \citealt{bro13}; $\kappa$\,CrB: \citealt{bon13}, including exozodiacal dust systems, e.g., around Vega: \citealt{abs06}, \citealt{def11}; Fomalhaut: \citealt{abs09}, \citealt{men13}, \citealt{leb13}; $\beta$\,Pic: \citealt{def12}). Some of these stars harbor (candidate) planetary companions at a similar distance from the star as the dust (i.e., HR8799: \citealt{mar08}, \citealt{mar10}; Fomalhaut: \citealt{kal08}; $\beta$\,Pic: \citealt{lag10}; our Solar system), possibly being responsible for the gap between the multiple components of the disk. This raises the question of whether such a companion might also exist around HIP\,17439. It is interesting to note that the configuration of the two rings as found by the present study closely resembles structures simulated by \citet{ert12b} (their model sequence~III), suggesting that a single, Jovian mass planet might be sufficient to explain the configuration of the belts in our best-fit model. However, it is important not to over interpret these similarities. On the one hand, the parameters found from our best-fit come with large uncertainties. On the other hand, it was not the goal of \citet{ert12b} to produce highly accurate structures suited for comparison with real observations, but to produce reasonable structures to be used for the subsequent analysis of their observability carried out. More detailed conclusions require better constraints on the configuration of the system as well as dedicated simulations of planet-disk interaction. A clear proof that planets are responsible for the multi-component structures in these systems would allow to study a new class of planets (large separation form the host star, possible low/intermediate mass or higher age compared to the planets discovered at such distance by direct imaging or hinted at by long term radial velocity trends) not accessible to any other observing technique available.

\begin{acknowledgements}
S.~Ertel and J.-C.~Augereau thank the French National Research Agency (ANR, contract ANR-2010 BLAN-0505-01, EXOZODI) and PNP-CNES for financial support. C.~Eiroa, J.~Maldonado, J.~P.~Marshall, and B.~Montesinos are partially supported by Spanish grant AYA~2011-26202. A.~V.~Krivov acknowledges support from the DFG, grant Kr~2164/10-1. T.~L\"ohne acknowledges support from the DFG, grant Lo~1715/1-1. R.~Liseau acknowledges the continued support by the Swedish National Space Board (SNSB). We thank the anonymous referee for valuable comments. S.~Ertel thanks K.~Ertel for general support.
\end{acknowledgements}

\bibliographystyle{aa}

\bibliography{bibtex}

\end{document}